\documentclass[amsmath,amssymb,reprint,aps,prl,showpacs,superscriptaddress,twocolumn,floatfix]{revtex4-1}

\usepackage{graphicx}
\usepackage[dvips]{color}

\begin{document}

\title{Engineering Negative Differential Conductance with the Cu(111) Surface State}
\author{B.\ W.\ Heinrich}
\altaffiliation{Institut f\"{u}r Experimentalphysik, Freie Universit\"{a}t Berlin, 14195 Berlin, Germany}
\author{ M.\ V.\ Rastei}
\author{D.-J.\ Choi}
\affiliation{Institut de Physique et Chimie des Mat\'{e}riaux de Strasbourg$\text{,}$ CNRS$\text{,}$ Universit\'{e} de Strasbourg, F-67034 Strasbourg, France}
\author{T.\ Frederiksen}
\affiliation{Donostia International Physics Center (DIPC), E-20018 Donostia-San Sebasti{\'a}n, Spain}
\author{L.\ Limot}
\email{limot@ipcms.unistra.fr}
\affiliation{Institut de Physique et Chimie des Mat\'{e}riaux de Strasbourg$\text{,}$ CNRS$\text{,}$ Universit\'{e} de Strasbourg, F-67034 Strasbourg, France}
\date{\today}

\begin{abstract}
Low-temperature scanning tunneling microscopy and spectroscopy are employed to investigate electron tunneling from a C$_{60}$-terminated tip into a Cu(111) surface. Tunneling between a C$_{60}$ orbital and the Shockley surface states of copper is shown to produce negative differential conductance (NDC) contrary to conventional expectations. NDC can be tuned through barrier thickness or C$_{60}$ orientation up to complete extinction. The orientation dependence of NDC is a result of a symmetry matching between the molecular tip and the surface states.
\end{abstract}

\pacs{68.37.Ef,72.80.Rj,73.20.At,73.63.-b,85.65.+h}

\maketitle 
Important advances have been made over the past decades in molecular electronics. Among these is the demonstration that molecules can perform controllable functions such as negative differential resistance or conductance (NDC) \cite{chen99,gaudioso00,tao06}.  First discovered in the Esaki diode \cite{esaki58}, NDC leads to regions in the $I-V$ curve where the current $I$ decreases (increases) with increasing (decreasing) voltage $V$. This fundamental property is nowadays exploited in CMOS devices for low-power memory, fast switches or oscillators. Early atomic-scale observations of NDC by scanning tunneling microscopy (STM) have been attributed to narrow energy states in tip and sample \cite{lyo89,bedrossian89}, in analogy with the resonant tunneling leading to NDC in semiconductors \cite{chang91}. In principle, resonant tunneling via molecular orbitals also leads to NDC in single molecules \cite{xue99,guisinger04,chen07}, but progress in this direction has been hindered by the lack of microscopic control over electrode and molecule status. 

It was shown in particular with single C$_{60}$ that NDC occurrence can be improved by narrowing the molecular levels through a reduced molecule-substrate coupling \cite{zeng00,franke08}, in order to exploit the bias-dependence of the transmission function \cite{grobis05,novaes11}. In this Letter, we introduce a different approach and report the occurrence of single-molecule NDC with a C$_{60}$-terminated tip. By attaching a molecule to the STM tip an increased control is gained over the entire tunnel junction \cite{schull09,gross11}. This method allows, unlike previous NDC studies, exploring the NDC occurrence with well-defined pristine metal surfaces serving as a conterelectrode.
Taking advantage of this setup, we demonstrate in stark contrast to conventional mechanisms that NDC can be produced by electron tunneling between a molecular orbital of the tip and a two-dimensional electron gas hosted by a copper surface \textemdash the Shockley surface states of Cu(111). In this calibrated setup, NDC may be tuned by varying the barrier thickness or by changing the C$_{60}$ orientation up to complete extinction. Our study demonstrates that molecular orbitals act as angular momentum filters to the tunneling process, leading, in particular, to NDC if accurately matched with the local orbital symmetry of the surface states. These findings should simplify NDC engineering at the atomic-scale.

The measurements were carried out with an STM operating at $4.6$~K in the $10^{-11}$~mbar range. After cleaning Cu(111) by sputter/anneal cycles, $0.15$~monolayers of C$_{60}$ were dosed onto the surface from a crucible containing a $99.9\%$ pure powder heated to $400^{\circ}$C. Single Au atoms were instead evaporated onto the cold surface by heating a high-purity gold wire. The differential conductance spectra, $dI/dV(V)$ were acquired via lock-in detection with a bias modulation ranging from $3$ to $10$~mV~rms at $670$~Hz (sample bias is measured with respect to the tip). The etched W tip employed was cleaned by sputter/anneal cycles and treated \textit{in vacuo} by soft indentations into the clean surface. The tip apex was therefore coated with copper.

\begin{figure}
\includegraphics[width=\columnwidth,clip=]{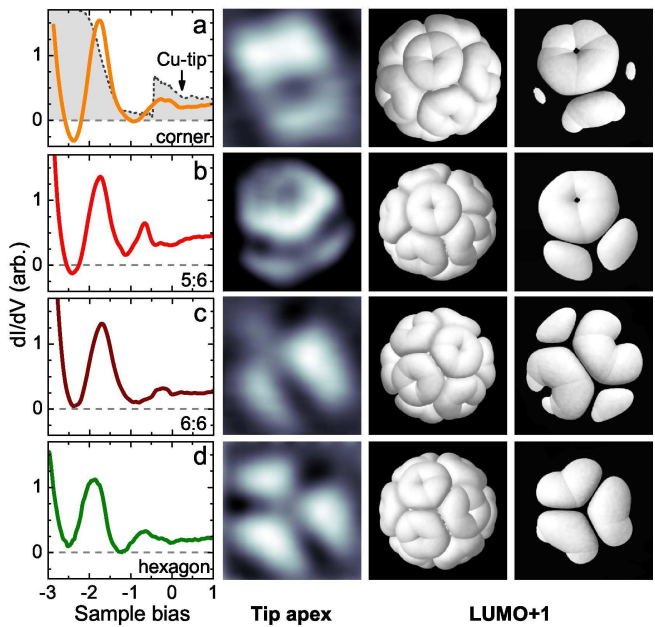}
  \caption{(color online). (a)\textemdash(d) $dI/dV$ over Cu(111) for various C$_{60}$ tips ($1.0$~V, $0.1$~nA). The orientation of C$_{60}$ at each tip apex is determined by constant-current images over a Au atom (sample bias: $-2.0$~V, current: $0.1$~nA, size: $12\times 12$~{\AA}$^2$). Each orientation is flanked by a H\"{u}ckel simulation of the LUMO+1 orbital, and for better visibility, by the same orbital cut along the $xy$ plane (the LUMO+1 orbital is localized on the pentagons forming the C$_{60}$ skeleton). Panel (a) present also a $dI/dV$ spectrum over Cu(111) acquired with a copper-coated W tip (dashed line, feedback loop opened at $-3.0$~V and $5$~nA).}
\label{fig1}
\end{figure}

Figure~\ref{fig1}(a) presents the electronic structure of the Cu(111) surface acquired with a metal tip (dashed line). Known spectroscopic features are evidenced \cite{kevan87}. The steplike feature at $-0.45$~V is associated to Shockley surface states, which are an experimental realization of a nearly two-dimensional electron gas; the marked upturn below $-0.9$~V is due to copper $d$-states present below the (111) band gap. To engineer NDC with a C$_{60}$ tip, we attach a single C$_{60}$ to the copper-coated tip by repeatedly bringing the tip into contact with the target molecule and pressing beyond contact \cite{schull09}. Figure~\ref{fig1}(a) presents a conductance spectrum acquired with a C$_{60}$ tip over Cu(111). The spectrum is nearly mirror symmetric to the spectrum acquired with a metallic tip positioned above a C$_{60}$ molecule [compare with Fig.~S2(b) in the supplementary material]. The lowest unoccupied molecular orbitals are located at negative bias (LUMO: $-0.5$~V, LUMO+1: $-1.9$~V), while the highest occupied orbitals are at positive bias (HOMO: $2.0$~V, not shown). The widths of the molecular states are typically $0.5$~V and compare well with a C$_{60}$ molecule adsorbed on a metal surface \cite{lu03,silien04,pai10}. Most importantly, NDC is detected below the LUMO+1. 

\begin{figure}
\includegraphics[width=\columnwidth,clip=]{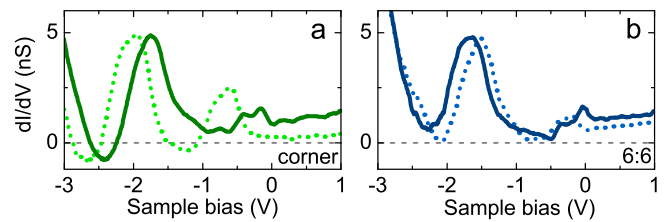}
  \caption{(color online). $dI/dV$ acquired above Cu(111) with two different tips ($1.0$~V, $1.0$~nA), having (a) same corner orientation for C$_{60}$ [as in Fig.~\ref{fig1}(a)], (b) same 6:6 orientation for C$_{60}$ [as in Fig.~\ref{fig1}(c)].}
\label{fig2}
\end{figure}

Among the numerous C$_{60}$ tips investigated, the ones producing NDC had a specific C$_{60}$ orientation relative to the copper surface. Figures~\ref{fig1}(a)\textemdash(d) summarize our findings. In order to image the tip apex, constant-current images of a Au atom were acquired \cite{schull09}. These images exhibit a molecular pattern matching the LUMO+1 for biases close to $-1.9$~V. The orientations reported in Fig.~\ref{fig1} correspond to a C$_{60}$ adsorbed on a carbon atom [labeled corner in Fig.\ref{fig1}(a)], on a pentagon-hexagon bond [5:6 in Fig.\ref{fig1}(b)], on a hexagon-hexagon bond [6:6 in Fig.\ref{fig1}(c)] and on a hexagonal ring [hexagon in Fig.\ref{fig1}(d)]. The apex C$_{60}$ was rotated by performing a tip contact with a Au atom \cite{neel08b}. While the corner and the 5:6 adsorptions favor NDC, other orientations tend to suppress it. We also found that the interface structure between the C$_{60}$ molecule and the copper tip has only a limited impact on NDC. Typical spectra acquired with different C$_{60}$-tips, but with the same corner or 6:6 orientation are presented in Figs.~\ref{fig2}(a) and~\ref{fig2}(b), respectively. Despite the C$_{60}$-copper interface changes in these tips, as indicated by the shift of the LUMO and LUMO+1 (of the order of $0.5$ and $0.2$~V, respectively) \cite{pai10}, the dominant mechanism leading to NDC is still the C$_{60}$ orientation.

Along with molecular orientation, counterelectrode selection is extremely crucial to NDC. In our setup, the counterelectrode can be conveniently chosen by positioning a C$_{60}$ tip above nanoscale objects supported by Cu(111) or by changing the metal surface. Several counterelectrodes were inspected, either having a featureless electronic structure over the energy range of interest such as a Au atom on Cu(111) or the Cu(100) surface, either presenting a marked electronic structure such as a cobalt islands on Cu(111) \cite{rastei07}, or a C$_{60}$ molecule on Cu(111). The $dI/dV$ spectra, presented as supplementary material, never exhibited NDC. The fact that NDC is instead observed on Au(111) as on Cu(111) \cite{schullprivate}, indicates that electron tunneling into the Shockley surface states is likely a key mechanism behind NDC.

\begin{figure}
\includegraphics[width=\columnwidth,clip=]{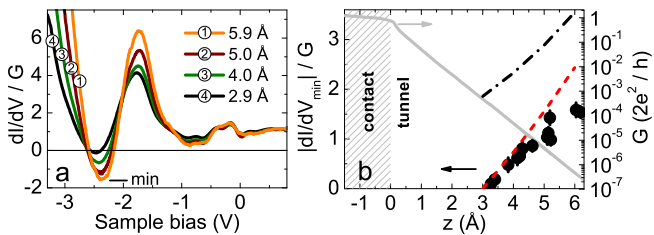}
  \caption{(color online). (a) $dI/dV$ spectra for various barrier thicknesses. The spectra are normalized by the conductance ($G=I/V$) at which the feedback loop was opened, respectively from top to bottom: $0.01$, $0.1$, $1$ and $10$~nS. (b) NDC intensity versus tip-sample distance (left vertical axis) and corresponding changes of $G$ (right vertical axis). From the tunneling regime, we extract an apparent-barrier height of $\Phi=5$~eV. The distance is estimated by taking $z=0$ for the contact conductance ($G$ vs $z$ curve acquired at $1$~V), the dashed area highlighting the contact regime. The NDC intensity corresponds to the minimum of the $dI/dV$ curves in (a). The lines correspond to the simulation detailed in the text:
a realistic DOS (dashed line) and a stepped DOS without $d$-bulk contribution (dash-dotted line).}
\label{fig3}
\end{figure}

For completeness, we investigated the influence of barrier thickness on NDC in the line of recent studies \cite{grobis05,franke08,tu08}. Figure~\ref{fig3}(a) presents a typical set of spectra acquired at various distances, while Fig.~\ref{fig3}(b) quantifies the changes in NDC with distance (noted $z$) for a collection of C$_{60}$-tips. Through a tip displacement it is possible to tune NDC and, unlike previous work, even switch it off well within the tunneling regime at a distance of $3$~{\AA} from surface. As established below, NDC extinction is a general property of our simplified setup not exclusive to C$_{60}$.

\begin{figure}
\includegraphics[width=\columnwidth,clip=]{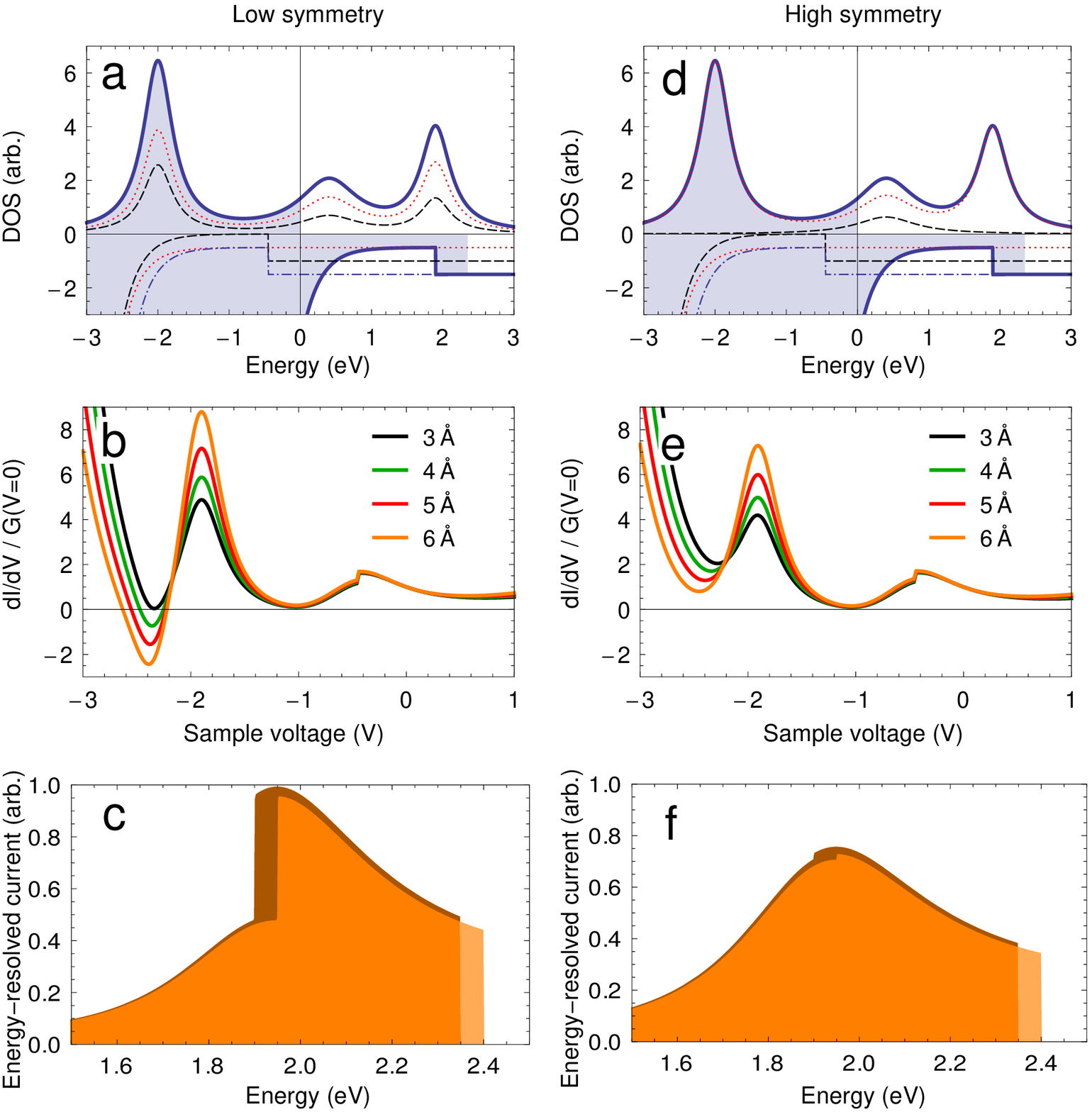}
  \caption{(color online). (a) DOS of low-symmetry C$_{60}$-tip (upper half-plane, thick line) and Cu(111) surface (lower half-plane, dash-dotted line: $V=0$~V, thick line: $V=-2.35$~V) used in the WKB simulation ($m=0$ and $m\neq 0$ components shown with dashed and dotted lines, respectively). Copper DOS: step function of unit amplitude centered at $-0.45$~eV in $\rho_s^{m=0}$ and a constant background set to $0.5$ in $\rho_s^{m\neq0}$ with low-energy upturn. C$_{60}$ DOS: three Lorentzian peaks mimicking the HOMO ($-2.0$~eV), LUMO ($0.4$~eV) and LUMO+1 ($1.9$~eV) with widths $0.5$~eV, $1.0$~eV, and $0.5$~eV, respectively, and amplitudes proportional to the free-molecule degeneracies (thick line). The occupied states are indicated by the shaded blue areas for the sample voltage $V=-2.35$ V where the Shockley surface state edge matches the LUMO+1 resonance. (b) Simulated $dI/dV$ for (a) via Eq.~(\ref{wkb}). (c) Energy-resolved current for $V=-2.35$V (dark orange) and $V=-2.40$~V (light orange) for $z=6$ {\AA}. (d)\textemdash(f) Similar to (a)\textemdash(c) but for a high-symmetry C$_{60}$ orientation where the  $m=0$ component is suppressed for HOMO and LUMO+1.}
\label{fig4}
\end{figure}

In the following we discuss the origin of the NDC with emphasis on the central role played by the Shockley surface states. To successfully engineer NDC, an accurate choice of the counterelectrode, of molecular orientation and of barrier thickness turns out to be crucial. To start, we adopt a WKB framework and focus on a junction comprising a sample metal surface and a C$_{60}$ tip with density of states (DOS) $\rho_s$ and $\rho_t$, respectively. The zero-temperature current $I$ as a function of sample voltage $V$ is then
\begin{equation}
I(V,z) \propto \int_{E_F-V}^{E_F}\sum_m\rho_s^m(E+V) \rho_t^m(E) T(E,V,z)dE,
\label{wkb}
\end{equation}
where $T(E,V,z)= \exp(-2\sqrt{2(\Phi-V/2-E)}z)$ is the transmission function in Hartree atomic units. The explicit summation over magnetic quantum numbers $m$ (projection of electron angular momentum on the common symmetry axis) excludes the extremely small probability for tunneling between states with different $m$ \cite{chen90,chen07}. For an accurate description of the tunneling current, we need just to distinguish between $m=0$ and $m\neq0$ in the DOS. This distinction is necessary as, near $\Gamma$, the $sp$-like Shockley surface states of Cu(111) only couple to the $m=0$ component of the tip states. To model the sample DOS, we thus take a stepped function centered at $-0.45$~eV for the surface states and a low-energy exponential upturn for the copper bulk $d$-states to represent $\rho_s^{m=0}$ [dashed line, negative half-plane, Fig.~\ref{fig4}(a)] and a constant background plus exponential upturn for $\rho_s^{m\neq0}$ [dotted line, negative half-plane, Fig.~\ref{fig4}(a)]. The total DOS (dash-dotted line) roughly mimics the electronic structure evidenced in Fig.~\ref{fig1}(a). For the C$_{60}$-tip we represent $\rho_t$ as a set of three Lorentzian peaks for the HOMO, LUMO, and LUMO+1 as described in the caption to Fig.~\ref{fig4}(a). The $m=0$ (dashed line) and $m\neq0$ (dotted line) components sum up (thick line) to amplitudes proportional to the free-molecule degeneracies. 

As shown in Fig.~\ref{fig4}(b), the corresponding simulated $dI/dV$ captures the essential experimental facts. There is indeed NDC in the spectrum just below the LUMO+1 down to distances close to $z=3$~{\AA}. The role of the Shockley surface states in the NDC is evident from the energy-resolved current shown in Fig.~\ref{fig4}(c): At a sample voltage $V=-2.35$ V the surface state onset matches the LUMO+1 resonance \cite{lyo89,bedrossian89}. Upon a slight increase in the voltage the surface state step feature moves \emph{away} from the molecular resonance. The associated loss in energy-resolved current (dark orange area) is only partially  compensated by the corresponding shift of the sample chemical potential (light orange area), the tunnel current effectively \emph{decreases} thus giving rise to NDC \cite{note}. The comparison to our experimental results is also satisfying as shown by the dashed line in Fig.~\ref{fig3}(b), where we also include the NDC produced by a stepped function (dash-dotted line) without the $d$-bulk upturn. The increased NDC intensity without the $d$-states highlights that they are detrimental to NDC. The peculiar decay reported in Fig.~\ref{fig3}(b) is therefore the result of a competition between opposite effects for $d$-states and Shockley surface states.

\begin{table*}
 \centering
\begin{tabular}{lccccc}\hline\hline
       & isolated & 5:6/corner & 6:6                      & hexagon  & pentagon\\
       & $I_h$    & $C_s$      & $C_{2v}$                 & $C_{3v}$ & $C_{5v}$\\\hline
HOMO   & $H_u$    & 2A$'$+3A$''$   & A$_1$+2A$_2$+B$_1$+B$_2$ & A$_2$+2E & A$_2$+E$_1$+E$_2$\\
LUMO   & $T_{1u}$ & 2A$'$+A$''$    & A$_1$+B$_1$+B$_2$        & A$_1$+E  & A$_1$+E$_1$\\
LUMO+1 & $T_{1g}$ & A$'$+2A$''$    & A$_2$+B$_1$+B$_2$        & A$_2$+E  & A$_2$+E$_1$\\\hline\hline
\end{tabular}
\caption{C$_{60}$ orbitals in different local symmetries imposed by molecular orientation with respect to an isotropic surface.}
\label{tab}
\end{table*} 

Another salient experimental finding is the impact of C$_{60}$ orientation on NDC, which was set aside in the above discussion. In Fig.~\ref{fig1} it was found that NDC occurs with C$_{60}$ in low-symmetry configurations (e.g., corner orientation) but disappears in high-symmetry configurations (e.g., hexagon orientation). Our NDC originates from the Shockley surface states which select $m=0$ tip states. This suggests that for the high-symmetry C$_{60}$ orientations the $m=0$ component of the LUMO+1 vanishes. Indeed such an effect can be rationalized by considering the transformation properties of the C$_{60}$ tip orbitals using group theory as follows.

For isolated C$_{60}$ molecules of icosahedral symmetry $I_h$ the HOMO is a fivefold degenerate state of $H_u$ symmetry while the LUMO and LUMO+1 are triply degenerate states of $T_{1u}$ and $T_{1g}$ symmetry, respectively. When the molecule is adsorbed on a surface these degeneracies will be reduced \cite{hands10}, and, importantly here, the orbital angular momentum components onto the common symmetry axis perpendicular to the surface will depend on the molecular orientation. In Tab.~\ref{tab} we list how the orbitals are split in different relevant local symmetry environments. For instance, a C$_{60}$ molecule adsorbed with a hexagon on an isotropic surface possesses a threefold rotation axis and can thus be considered as belonging to the $C_{3v}$ group. As a consequence it can be shown that the $T_{1g}$ LUMO+1 splits into A$_2$+E representations \cite{carter}. However, from the character table of $C_{3v}$ it is evident that $s$, $p_z$, and $d_{z^2}$ orbitals do not transform according to those representations. In other words, the LUMO+1 does not have a $m=0$ component for the hexagon orientation. Contrarily, a C$_{60}$ molecule adsorbed on a 5:6 bond or a corner atom only possesses bilateral symmetry as described by $C_s$, and the LUMO+1 is split into A$'$+2A$''$. As $s$, $p_z$, and $d_{z^2}$ \emph{do} transform in $C_s$ according to  those representations the LUMO+1 consequently contains a $m=0$ component in these low-symmetry configurations. 

More generally, based on Tab.~\ref{tab} we can conclude that tunneling into the surface state is symmetry forbidden for orbitals which do not transform as the totally symmetric representation (A$'$/A$_1$), $i.\ e.$, the HOMO for pentagon and hexagons orientations and the LUMO+1 for 6:6, pentagon, and hexagon orientations. Note also that there are no such symmetry constraints for the LUMO orbital. To simulate the $dI/dV$ recorded with C$_{60}$ tips with high-symmetry orientation we therefore eliminated the $\rho_t^{m=0}$ component in HOMO and LUMO+1 [Fig.~\ref{fig4}(d)]. As shown in Fig.~\ref{fig4}(e)\textemdash(f), indeed this leads to the disappearance of NDC as was observed experimentally [Fig.~\ref{fig1}(c)\textemdash(d)].

To summarize, we have shown that electron tunneling between a well calibrated molecule and a two-dimensional electron gas produces a tunable NDC. By controlling both tip and sample states and accounting for their symmetries, we demonstrated the concept of local orbital symmetry matching \cite{chen07}. Barrier thickness or symmetry matching between the molecule and the counterelectrode can be used to switch the NDC on and off.  As dispersive two-dimensional electronic states are widely available in metals and semiconductors, and moreover may be controllably modified through artificial nanostructures or molecular adlayers, our findings should simplify NDC engineering in atomic-scale metal-organic junctions.

\begin{acknowledgments}
We thank G.\ Schull, J.\ I.\ Pascual, W.\ A.\ Hofer, and D.\ S\'anchez-Portal for fruitful discussions. 
\end{acknowledgments}


\begin{thebibliography}{nnnyys}

\bibitem{chen99} J. Chen, M. A. Reed, A. M. Rawlett, and J. M. Tour, Science \textbf{286}, 1550 (1999).

\bibitem{gaudioso00} J. Gaudioso, L. J. Lauhon, and W. Ho, Phys. Rev. Lett. \textbf{85}, 1918 (2000).

\bibitem{tao06} N. J. Tao, Nat. Nanotech. \textbf{1}, 173 (2006).

\bibitem{esaki58} L. Esaki, Phys. Rev. \textbf{109}, 603 (1958).

\bibitem{lyo89} I.-W. Lyo and P. Avouris, Science \textbf{245}, 1369 (1989). 

\bibitem{bedrossian89} P. Bedrossian, D. M. Chen, K. Mortensen, and J. A. Golovchenko, Nature \textbf{342}, 258 (1989). 

\bibitem{chang91} L. Chang, E. Mendez, and C. Tejedor, eds., \textit{Resonant Tunneling in Semiconductors} (Plenum, New York, 1991).

\bibitem{xue99} Y. Xue \textit{et al.}, Phys. Rev. B \textbf{59}, R7852 (1999).

\bibitem{guisinger04} N. P. Guisinger \textit{et al.}, Nano Lett. \textbf{4}, 55 (2004).

\bibitem{chen07} L. Chen \textit{et al.}, Phys. Rev. Lett. \textbf{99}, 146803 (2007).

\bibitem{zeng00} C. Zeng \textit{et al.}, Appl. Phys. Lett. \textbf{77}, 3595 (2000).

\bibitem{franke08} K. J. Franke \textit{et al.}, Phys. Rev. Lett. \textbf{100}, 036807 (2008).

\bibitem{grobis05} M. Grobis, A. Wachowiak, R. Yamachika, and M. F. Crommie, Appl. Phys. Lett. \textbf{86}, 204102 (2005).

\bibitem{novaes11} F. D. Novaes \textit{et al.}, arXiv:1101.3714v1.

\bibitem{schull09} G. Schull, T. Frederiksen, M. Brandbyge, and R. Berndt, Phys. Rev. Lett. \textbf{103}, 206803 (2009).

\bibitem{gross11} L. Gross \textit{et al.}, Phys. Rev. Lett. \textbf{107}, 086101 (2011).

\bibitem{kevan87} S. D. Kevan and R. H. Gaylord  Phys. Rev. B \textbf{36}, 5809 (1987). 

\bibitem{neel08b} N. N{\'e}el, L. Limot, J. Kr{\"o}ger, and R. Berndt, Phys. Rev. B \textbf{77}, 125431 (2008).

\bibitem{lu03} X. Lu \textit{et al.}, Phys. Rev. Lett. \textbf{90}, 096802 (2003).

\bibitem{silien04} C. Silien, N. A. Pradhan, W. Ho, and P. A. Thiry, Phys. Rev. B \textbf{69}, 115434 (2004).

\bibitem{pai10} W. W. Pai \textit{et al.}, Phys. Rev. Lett. \textbf{104}, 036103 (2010).

\bibitem{rastei07} M.\ V.\ Rastei \textit{et al.}, Phys. Rev. Lett. \textbf{99} 246102 (2007).

\bibitem{schullprivate} G. Schull and R. Berndt (private communication).

\bibitem{tu08} X. W. Tu, G. R. Mikaelian, and W. Ho, Phys. Rev. Lett. \textbf{100}, 126807 (2008).

\bibitem{chen90} C. J. Chen, Phys. Rev. B \textbf{42}, 8841 (1990).

\bibitem{note} Simulations also show that NDC is favoured when the energy position of the surface-state onset is comparable to the width of the LUMO+1, which is indeed the case in the present setup.

\bibitem{hands10} I. D. Hands, J. L. Dunn, and C. A. Bates, Phys. Rev. B \textbf{81}, 205440 (2010).

\bibitem{carter} R. L. Carter, \textit{Molecular Symmetry and Group Theory} (John Wiley \& Sons, Inc. 1998).

\end{thebibliography}
\end{document}